\documentclass[a4paper,fleqn,usenatbib]{mnras}

\usepackage{newtxtext,newtxmath}
\usepackage[T1]{fontenc}
\usepackage{ae,aecompl}

\usepackage{graphicx}	% Including figure files
\usepackage{amsmath}	% Advanced maths commands
\usepackage{amssymb}	% Extra maths symbols
\usepackage{xcolor}

\newcommand{\cmc}{\,cm$^{-3}$}  % per cm-cubed
\newcommand{\HII}{H\,{\sc ii}} % 

\title[Hydrogen RRLs]{The influence of continuum radiation fields on hydrogen 
radio recombination lines}

\author[A. Prozesky and D. P. Smits]{
Andri Prozesky,$^{1}$\thanks{E-mail: prozea@unisa.ac.za (AP)}
and Derck P. Smits$^{1}$
\\
$^{1}$Department of Mathematical Sciences, University of South Africa, Private 
Bag X6, Florida 1709, South Africa\\
}

\date{Accepted XXX. Received YYY; in original form ZZZ}

\pubyear{2017}

\begin{document}
\label{firstpage}
\pagerange{\pageref{firstpage}--\pageref{lastpage}}
\maketitle

\begin{abstract}
Calculations of hydrogen departure coefficients using a model with the angular 
momentum quantum levels resolved that includes the effects of external radiation 
fields are presented. The stimulating processes are important at radio frequencies 
and can influence level populations. New numerical techniques with a solid 
mathematical basis have been incorporated into the model to ensure convergence of 
the solution. Our results differ from previous results by up to 20 per cent. A 
direct solver with a similar accuracy but more efficient than the iterative method 
is used to evaluate the influence of continuum radiation on the hydrogen population 
structure. The effects on departure coefficients of continuum radiation from dust, 
the cosmic microwave background, the stellar ionising radiation, and free-free 
radiation are quantified. Tables of emission and absorption coefficients for 
interpreting observed radio recombination lines are provided.
\end{abstract}

\begin{keywords}
atomic processes -- line: formation -- \HII\ regions -- methods: numerical -- ISM: 
atoms -- radio lines: ISM \\
\end{keywords}

%%%%%%%%%%%%%%%%%%%%%%%%%%%%%%%%%%%%%%%%%%%%%%%%%%

%%%%%%%%%%%%%%%%% BODY OF PAPER %%%%%%%%%%%%%%%%%%

%%%%%%%%%%%%%%%%% INTRODUCTION %%%%%%%%%%%%%%%%%%%%%

\section{Introduction}
\defcitealias{1995Storey}{SH95}

\begin{figure}
	\includegraphics[width=\columnwidth]{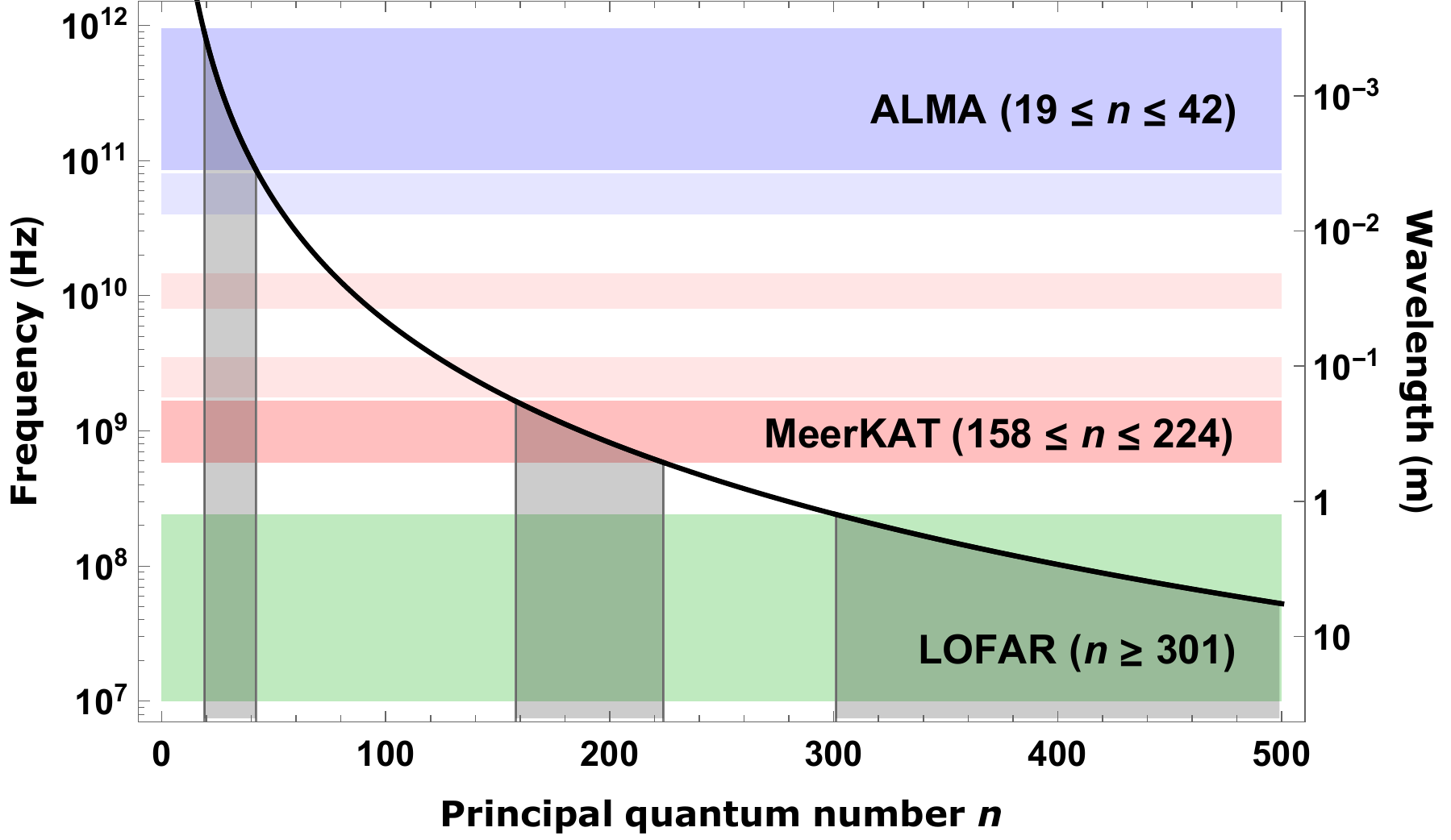}
    \caption{The black line shows the frequencies (wavelengths) of H$n\alpha$ 
transitions as a function of $n$ for hydrogen. The horizontal shaded bands depict 
the operating frequency bands of ALMA (blue), MeerKAT (red) and LOFAR (green). 
The lighter shaded bands are planned additions to the telescopes, but are not 
operational yet. The vertical grey shaded areas indicate the ranges of principal 
quantum number for which the H$n\alpha$ transitions will technically fall into 
the observing frequencies of the respective telescopes. A colour version of this 
graph is available in the electronic copy of this article.}
    \label{fig:alphatrans}
\end{figure}

Emission lines from gaseous nebulae contain a wealth of information on the 
conditions in the plasma, but to interpret observed spectra theoretical models of 
line fluxes are required. The construction of new radio telescopes such as MeerKAT, 
ALMA and LOFAR provides new opportunities to study emission lines with improved 
sensitivity at low frequencies. Fig.~\ref{fig:alphatrans} shows the H$n\alpha$ 
transitions (hydrogen transitions between principal quantum numbers $n+1 
\rightarrow n$) that will be observable with each of these telescopes. Low 
frequency ($\leq 1.4$\,GHz) detections of hydrogen recombination lines are rare 
\citep{2002Anantharamaiah}, but carbon lines are commonly detected at these 
frequencies. Transitions in carbon produce lines with different frequencies to 
hydrogen, but reliable calculations of lines with $n \gtrsim 300$ are needed to 
interpret results from insturments such as LOFAR. 

\citet{1966goldberg} showed that at low frequencies where stimulated processes 
are important the accuracy of the calculation of atomic level populations has a 
substantial influence on the theoretical intensities of recombination lines. This 
sentiment is echoed 50 years later by \citet{2017SanchezContreras} who also point 
out that there is disagreement between the results of various authors without one 
set of values clearly being the correct one. The need for accurate values is even 
more important now that radio telescopes with high sensitivity are being constructed. 

\citet{1976Burgess} included stimulated emission and absorption terms for the 
bound-bound and bound-free transitions in an $n$-model that went up to principal 
quantum number $n = 500$. They found that stimulated processes can have a 
significant effect on the values of the departure coefficients. This work was 
expanded by \citet{1977Summers} who resolved the angular momentum states for 
levels with $n \leq 35$, but only considered how the $^1$S, $^2$S and $^2$P levels 
of hydrogen were affected by a stellar radiation field. \citet{1977CoPhC..13...39B} 
and \citet{1979Salem} published their programme and tables based on the $n$-model 
of \citet{1970Brocklehurst} that gave departure coefficients for $50 \leq n \leq 
300$. The programme included updated collisional cross sections from \citet{1976Gee}, 
collisions due to protons, as well as radiative processes involving an external 
field, but did not consider angular momentum changing collisions. The programme 
was modified by \citet{1990Walmsley} to generate results down to $n =20$.

The results of \citet{1995Storey} (hereafter \citetalias{1995Storey}) are 
considered to be the definitive values for optical/IR recombination lines, but 
stimulated processes are not included in their model.  However, the values of 
\citetalias{1995Storey} are being used to study low frequency lines  
\citep[see for example][]{2006Fujiyoshi,2017Bendo,2017SanchezContreras}. In this 
paper we present results using a model that includes stimulated and absorption 
processes in the bound-bound and bound-free transitions in an $nl$-model, and use 
updated numerical methods.  Our results are compared to those of 
\citetalias{1995Storey}. The current calculated values differ significantly from 
their results for levels with principal quantum number $n\gtrsim30$.

In section~\ref{sec:model} we describe our model and calculational details. A 
stopping criterion for an iterative method of solution is derived in 
section~\ref{sec:stopping_criterion}, and the results of including this in our 
calculations are compared with those of \citetalias{1995Storey}. The use of a 
direct solver is also discussed here. The stimulating effects of continuum 
radiation fields within a nebula are considered in section~\ref{sec:cont_fields}. 
Finally, the main conclusions of this article are outlined in 
section~\ref{sec:conclusions}. Details of the atomic calculations used in our 
model are presented in the appendix.

%%%%%%%%%%%%%%%%% THE MODEL %%%%%%%%%%%%%%%%%%%%%

\section{The model}
\label{sec:model}

To determine theoretical line intensities, the electron populations $N_{nl}$ of 
all bound states need to be calculated. The level populations $N_{nl}$ are described 
by statistical equilibrium which requires that the total rate of all transitions 
into any particular level must equal the total rate of the transitions out of 
that level. 

\citet{1937Menzel} introduced a correction factor, denoted by $b_{nl}$, to 
compensate for the degree of departure from local thermodynamic equilibrium (LTE) 
of the level population from the LTE value $N_{nl}^*$ so that 
\begin{equation}
b_{nl}=\frac{N_{nl}}{N_{nl}^*}\,.
\end{equation}
A departure coefficient $b_{nl}$ that is equal to unity indicates that the level 
$nl$ is in LTE with the electron gas. In this scheme, the Saha-Boltzmann equation 
for hydrogen becomes 
\begin{equation}
N_{nl}=b_{nl}N_\mathrm{e} N_\mathrm{p} \left( \frac{h^2}{2\pi m_\mathrm{e}k_\mathrm{B}
T_\mathrm{e}}\right)^{3/2} (2l+1) 
\exp\left(\frac{\chi_{nl}}{k_\mathrm{B}T_\mathrm{e}}\right)\,,\label{eq:sahaboltznl}
\end{equation}
where $N_\mathrm{e}$ and $N_\mathrm{p}$ are the number density of electrons and 
protons respectively, $h$ is Planck's constant, $m_\mathrm{e}$ is the electron 
mass, $k_\mathrm{B}$ is the Boltzmann constant and $T_\mathrm{e}$ is the kinetic 
temperature of the free electron gas. The ionisation potential of level $nl$ for 
hydrogen is given by $\chi_{nl}=R_\mathrm{H}hc/n^2$ where $R_\mathrm{H}$ is the 
Rydberg constant for hydrogen and $c$ the speed of light in a vacuum. 

Collisional processes within the plasma become more efficient than their radiative 
counterparts as the principal quantum number $n$ of the level increases. Therefore, 
for each set of physical conditions there will be an $n=n^*$ where the collisional 
processes dominate completely and set up Boltzmann distributions with temperature 
$T_\mathrm{e}$ among the levels so that $b_{nl}=1$ for $n\geq n^*.$

\citet{1966goldberg} introduced an amplification factor $\beta_{n,n'}$ with $n>n'$
that is given by
\begin{equation}
\beta_{n,n'} = \frac{\left(1-\frac{b_{n'}}{b_n} e^{-h\nu/k_\mathrm{B}T_\mathrm{e}} 
\right)}{\left(1-e^{-h\nu/k_\mathrm{B}T_\mathrm{e}}\right)}
\label{eq:betadef}
\end{equation}
where $\nu$, the frequency associated with the $n$-$n'$ transition, is given by
\begin{equation}
\nu=R_\mathrm{H} c \left( \frac{1}{n^{'2}}-\frac{1}{n^2} \right)\,.
\end{equation}

The amplification factors give the departure of the ratios of level populations 
from what is expected in LTE, as opposed to the departure coefficients that 
represent the departure of individual level populations from LTE. A value of 
$\beta_{n,n'}<0$ indicates a population inversion between levels $n$ and 
$n'$, with an increased amount of inversion for smaller values of $\beta_{n,n'}$. 
Therefore,  stimulated emission will be important for levels for which 
$\beta_{n,n'} \ll 0$. An illustrative discussion of the amplification factor can 
be found in \citet{1996Strelnitski}.

\citet{1938BakerMenzel} introduced two simple assumptions which are referred to 
as Case A and Case B. For Case A, the nebula is taken to be optically thin to all 
line radiation. For Case B, it is assumed that all photons produced by Lyman 
transitions are optically thick and are absorbed close to the point where they 
are emitted (this is called the on-the-spot approximation). From a calculational 
perspective, this means that all transitions to the $n = 1$ level are 
ignored. \citet{1962Osterbrock} concluded that Case B is a good quantitative 
approximation for nebular conditions.

Substituting the Saha-Boltzmann equation~(\ref{eq:sahaboltznl}) into the rate 
equation of all processes due to statistical balance yields
\begin{align}
 &  \left(\frac{h^2}{2\pi m_\mathrm{e}k_\mathrm{B}T_\mathrm{e}}\right)^{-3/2}
\left(\alpha_{nl}^\mathrm{r} + \alpha_{nl}^\mathrm{s} + N_\mathrm{e}C_{i,nl}\right) 
\ + \nonumber\\
 &   \underset{m>n}{\sum}\;\underset{l'=l\pm1}{\sum}b_{ml'}(2l'+1) 
e^{\chi_{ml'}/k_\mathrm{B}T_\mathrm{e}} \left( A_{ml',\,nl} + B_{ml',\,nl}J_{\nu} 
+ N_\mathrm{e}C_{ml',\,nl} \right) \nonumber\\
 & +  \underset{k<n}{\sum}\;\underset{l^{''}=l\pm1}{\sum}b_{kl^{''}}(2 l^{''} +1) 
e^{\chi_{kl^{''}}/k_\mathrm{B}T_\mathrm{e}} \left(B_{kl^{''},\,nl}J_{\nu} + 
N_\mathrm{e}C_{kl^{''},\,nl} \right)\nonumber\\
 &    +\underset{l'=l\pm1}{\sum}b_{nl'}(2l'+1)e^{\chi_{nl'}/k_\mathrm{B}
T_\mathrm{e}}\sum_{q} N_qC^q_{nl',\,nl}\nonumber\\
 & = \ b_{nl}(2l+1)e^{\chi_{nl}/k_\mathrm{B}T_\mathrm{e}}\left[\alpha_{nl}^\mathrm{p} 
+ N_\mathrm{e}C_{nl,i} \right.\nonumber\\
&  \left.+\underset{k<n}{\sum}\;\underset{l^{''}=l\pm1}{\sum} \left(A_{nl,\,kl^{''}} 
+ B_{nl,\,kl^{''}}J_{\nu} + N_\mathrm{e}C_{nl\,kl^{''}}\right)
\right. \nonumber\\
 &   \left.+\underset{m>n}{\sum}\;\underset{l'=l\pm1}{\sum}\left(B_{nl,\,ml'} 
J_{\nu} + N_\mathrm{e}C_{nl,\,ml'}\right) \ + \underset{l'=l\pm1}{\sum}\sum_{q}N_q 
C^q_{nl,\,nl'} \right]\,.
\label{eq:saharatenl}
\end{align}

The left-hand side contains all processes that populate level $nl$. The terms 
represent radiative recombination ($\alpha_{nl}^\mathrm{r}$), stimulated recombination 
($\alpha_{nl}^\mathrm{s}$), three-body recombination ($C_{i,nl}$), spontaneous emission 
($A_{nl,\,n'l'}$), stimulated emission ($B_{nl,\,n'l'}$), collisional de-excitation 
($C_{nl,\,n'l'}$), absorption ($B_{n'l',\,nl}$), collisional excitation 
($C_{n'l',\,nl}$) and elastic collisions ($C^q_{nl,\,nl'}$). The mean intensity 
of incident radiation fields is given by $J_\nu$.

The right-hand side includes all processes that depopulate level $nl$. The terms 
represent photoionisation ($\alpha_{nl}^\mathrm{p}$), collisional ionisation 
($C_{nl,i}$), spontaneous emission, stimulated emission, collisional de-excitation, 
collisional excitation, absorption and elastic collisions, respectively.

The $N_q$ represent the number densities of the different species interacting via 
elastic collisions with the bound electrons. In this work, electrons, protons and 
He$^+$ ions are taken to induce these collisions, with the proton number density 
$N_\mathrm{p} = 0.909 N_\mathrm{e}$ and the He$^+$ number density 
$N_{\mathrm{He}^+} = 0.090 N_\mathrm{e}$. These are the same values used by 
\citetalias{1995Storey}.

Equation~(\ref{eq:saharatenl}) represents a system of linear equations in the 
$b_{nl}$ values. Therefore, it can be written in matrix form as
\begin{equation}
\mathbfss{A}\mathbf{\cdot}\mathbf{b}=\mathbf{y}\label{eq:nlsys}
\end{equation}
where $\mathbf{b}$ is a vector with the $b_{nl}$ values as entries. The diagonal 
entries of the matrix $\mathbfss{A}$ represent the processes depopulating a level 
$nl$ and the off-diagonal entries in a given row are the bound-bound processes 
that are populating that level. Only dipole transitions are considered in this 
work which results in $\mathbfss{A}$ being sparse, i.e.~most of the entries in 
the $\mathbfss{A}$ matrix are equal to zero. The vector $\mathbf{y}$ contains the 
first term of equation~(\ref{eq:saharatenl}) as well as the populating contributions 
of levels with $n>n_{\mathrm c}$.

The values of the $b_{nl}$'s for a specific set of conditions are found in a two 
step process. First, a set of equations for which the angular momentum states are 
not resolved are solved to obtain values for $b_n$. The values produced by this 
method are valid for $n>n_\mathrm{c}$ where the equation~(\ref{eq:statpop}) holds. 
This is referred to as the $n$-model. 

Next, equations~(\ref{eq:saharatenl}) are solved explicitly for $n\leq n_\mathrm{c}$ 
to obtain the $b_{nl}$ values for this range.  This part is referred to as the 
$nl$-model and the $b_n$'s for this range are obtained via equations~(\ref{eq:bnsum}). 
Sections~\ref{sec:nmodel} and \ref{sec:nlmodel} provide details regarding the the 
two parts of the model and calculational details of the individual rates in 
equation~(\ref{eq:saharatenl}) are given in Appendix~\ref{sec:calcs}.

\subsection{The $n$-model}
\label{sec:nmodel}

The approach of \citet{1970Brocklehurst} was followed for the $n$-model. In 
addition to the processes in the Brocklehurst model, our $n$-model includes 
stimulated emission and absorption terms. In principle, an isolated atom has an 
infinite number of energy levels. To make the mathematics computationally viable, 
an upper cut-off $n_{\text{max}}$ was introduced for the highest $n$ level for 
which the rate equations were solved explicitly. The contributions to the sums 
above $n_{\text{max}}$ were converted into an integral using the trapezoidal rule. 
This integral was then approximated using a 20-point Gaussian quadrature. The 
remaining rate equations were cast into matrix form.

Because the departure coefficients vary smoothly and slowly with $n$, the Lagrange 
interpolation technique of \citet{1969Burgess} was employed to reduce the number 
of equations to be solved five fold. The condensed rate equations were solved by 
direct Gaussian elimination with the use of partial pivoting. The resulting 
condensed vector containing the $b_n$ values was interpolated to give the full 
set of values. The $b_n$ values found using this method compared very well with 
the results of solving the full system of equations.

\subsection{The $nl$-model}
\label{sec:nlmodel}

To obtain the departure coefficients for  $n\leq n_\mathrm{c}$, the rate 
equations~(\ref{eq:saharatenl}) have to be solved simultaneously. The main 
difference between the $b_n$'s from the $n$-model and the ones obtained from the 
$nl$-model and equation~(\ref{eq:bnsum}) is the inclusion of the elastic collisions 
between bound electrons and free particles. The elastic collisions' effects are 
important for the populations of mid- to high-energy levels where the $b_n$'s 
calculated with the $nl$-model can differ significantly from those of the $n$-model.

The semi-classical impact-parameter formulation developed by 
\citet{1964PengellySeaton} for the rates of angular momentum changing collisions 
has been considered definitive for many decades. Recently, \citet{2012Vrinceanu} 
presented updated formulae for these transition rates for both the quantum and 
semi-classical case. \citet{2016Guzman} did an in-depth analysis of the two 
approaches and concluded that the analytic equations presented in 
\citet{1964PengellySeaton} are much faster to compute and agree very well with 
the exact quantum mechanical probabilities of \citet{2012Vrinceanu}. This supports  
an earlier conclusion of \citet{2015StoreySochi}. In this work, the formalism of 
\citet{1964PengellySeaton} was followed with the modification of \citet{2016Guzman} 
to get the partial rates directly. Details are given in section~\ref{sec:elascol}.

There will be a $n=n_\mathrm{c}<n^*$ where the collisional processes will be much 
faster than the radiative processes, but radiative effects are still evident. For 
$n_\mathrm{c}< n<n^*$, the angular momentum states are populated statistically 
according to
\begin{equation}
N_{nl}=\frac{2l+1}{n^2}N_n\,.
\label{eq:statpop}
\end{equation}
The departure coefficient, $b_n$, that represents the departure from LTE for an 
energy level $n$ is defined as the weighted sum of the $b_{nl}$'s 
\begin{equation}
b_n = \frac{1}{n^2} \sum_{l=0}^{n-1} (2l+1) b_{nl} \,.
\label{eq:bnsum}
\end{equation}
Therefore, it follows that $b_n = b_{nl}$ for $n > n_\mathrm{c}$.

Equation~(\ref{eq:saharatenl}) represents a total of $n_\mathrm{c}(n_\mathrm{c}+1)/2$ 
equations that have to be solved simultaneously. This value is generally ${\sim}10^4$. 

The populating contributions of the first $20$ energy levels beyond $n_\mathrm{c}$ 
are calculated explicitly and incorporated into the vector $\mathbf{y}$ in 
equation~(\ref{eq:nlsys}). The populating effects of levels $n>n_\mathrm{c}+20$ 
are approximated with a 20-point Gaussian quadrature in the same way as the 
contributions of $n>n_{\text{max}}$ are handled in the $n$-model and added to 
$\mathbf{y}$. The depopulating effects of levels with $n_\mathrm{c}<n\leq 
n_\mathrm{c}+20$ and $n>n_\mathrm{c}+20$ are treated similarly, but added to the 
total depopulation rate for each level contained in the diagonal of the matrix 
$\mathbfss{A}$.

The level $n_\mathrm{c}$ is determined empirically to ensure the results of the 
$nl$-model are used up to a sufficiently high principal quantum number so that 
they match the results of the $n$-model to four significant digits. The level 
above which equation~(\ref{eq:bnsum}) is valid was found to be weakly dependent 
on temperature and is calculated using
\begin{equation}
n_\mathrm{c} = 350 - 15\ln(N_\mathrm{e})
\end{equation}
and rounding up to the nearest multiple of $5$.

%%%%%%%%%%%%%%%%% STOPPING CRITERION %%%%%%%%%%%%%%%%%%%%%

\section{Stopping criterion}
\label{sec:stopping_criterion}

\subsection{Background}

Because the values of $b_{nl}$ do not vary smoothly, the matrix condensation 
technique used in the $n$-model cannot be employed at this step. The standard 
method of solving the large number of simultaneous equations in the $nl$-model is 
to use an iterative method \citep[e.g.][]{1971Brocklehurst,1991Smits,1995Storey,
2017Salgado}.  An important aspect of an iterative solution is a test to indicate 
when convergence of the departure coefficients has been achieved, and to stop the 
iterations. In most previous work, the stopping criterion has not been explicitly 
stated, except for \citet{2017Salgado}, who terminate their iterative procedure 
when the difference between two successive iterations is less than 1 per cent. 

We show here that such a stopping criterion is not appropriate and can stop the 
iterative process before the values of the departure coefficients have converged 
to values within a known error. If the iterative procedure is making small 
corrections to the $b_{nl}$'s, the convergence rate can be slow, with the result
that a stopping criterion such as the one used by \citet{2017Salgado} can signal 
convergence has occured. Many more iterations can be required before convergence 
is reached. These tiny corrections can accumulate to a significant number, as will 
be shown below. For the conditions considered in this work, the matrix 
$\mathbfss{A}$ has a condtion number $\kappa(\mathbfss{A}) \sim 10^7$ which is 
very large and means the matrix is ill-conditioned, which results in a slow 
convergence rate \citep{Ill16}.

\subsection{Derivation}

To derive a stopping criterion, we define the following quantities and notation. 
Let $\mathbf{b}^{(i)}$ be the vector containing the departure coefficients as 
entries after $i$ iterations. The residual $\mathbf{r}^{(i)}$ after $i$ iterations 
is given by 
\begin{equation}
\mathbf{r}^{(i)}=\mathbfss{A}\mathbf{\cdot}\mathbf{b}^{(i)}-\mathbf{y}
\label{eq:residual}
\end{equation}
and provides an indication of the quality of $\mathbf{b}^{(i)}$. The vector of 
errors $\mathbf{e}^{(i)}$ after $i$ iterations is the difference between 
$\mathbf{b}^{(i)}$ and the true solution $\mathbf{b}$, i.e. 
\begin{equation}
\mathbf{e}^{(i)}=\mathbf{b}^{(i)}-\mathbf{b}\,.\label{eq:error}
\end{equation}

Of course, $\mathbf{b}$ is generally unknown and therefore $\mathbf{e}^{(i)}$ 
cannot be calculated directly. However, if an upper bound for the error can be 
defined then a stopping criterion can be constructed such that the iterations will 
only stop after it is guaranteed that the errors are smaller than some predefined 
number. 

A norm of a vector or matrix, indicated using double bars $\| \cdot \|$, is a 
non-negative number that gives a measure of the magnitude of the vector or matrix. 
There are various types of norms that can be defined, all of which obey a specified 
set of properties. In particular, for a matrix $\mathbfss{M}$ and vector 
$\mathbf{v}$ the inequality
\begin{equation}
\| \mathbfss{M}\mathbf{v} \|   \leq \| \mathbfss{M} \| \| 
\mathbf{v} \| \label{eq:submult}
\end{equation}
will hold for any consistent pair of vector and matrix norms. Note 
that the quantity called the condition number mentioned above is defined as
\begin{equation}
\kappa(\mathbfss{M}) = \| \mathbfss{M}^{-1} \| \ \| \mathbfss{M} \|\, . \label{eq:conno}
\end{equation}
If $\kappa(\mathbfss{M}) \gg 1$ then the matrix is called ill-conditioned.

Using the tools developed above, an appropriate stopping criterion for an iterative 
solution to the set of equations represented by equation~(\ref{eq:nlsys}) can now 
be derived. Substituting equation~(\ref{eq:nlsys}) into equation~(\ref{eq:error}) 
and using equation~(\ref{eq:residual}) gives
\begin{align}
\mathbf{e}^{(i)} & = \mathbf{b}^{(i)}-\mathbfss{A}^{-1}\mathbf{y}\nonumber\\
 & =\mathbfss{A}^{-1}(\mathbfss{A}\mathbf{\cdot}\mathbf{b}^{(i)}-\mathbf{y})\nonumber\\
 & =\mathbfss{A}^{-1}\mathbf{r}^{(i)}.
\end{align}
Taking the norm on both sides and using the submultiplicative property given in 
equation~(\ref{eq:submult}) leads to
\begin{equation}
\| \mathbf{e}^{(i)}\| = \| \mathbfss{A}^{-1}\mathbf{r}^{(i)}\| \leq\|
\mathbfss{A}^{-1}\| \| \mathbf{r}^{(i)}\| \,.
\end{equation}

Therefore, stopping the iterative procedure only after
\begin{equation}
\| \mathbf{r}^{(i)}\| \leq\epsilon\mathbf{\cdot}\frac{\| \mathbf{b}^{(i)}\| }{\| 
\mathbfss{A}^{-1}\| }\label{eq:stop}
\end{equation}
will guarantee that the relative error is smaller than some predefined tolerance 
$\epsilon\ll1$. That is,  equation~(\ref{eq:stop}) implies
\begin{equation}
\frac{\| \mathbf{e}^{(i)}\| }{\| \mathbf{b}^{(i)}\| }\leq\frac{\| 
\mathbfss{A}^{-1}\| \| \mathbf{r}^{(i)}\| }{\| \mathbf{b}^{(i)}\| }\leq\epsilon\,.
\label{eq:conv_criterion}
\end{equation}

Equation~(\ref{eq:conv_criterion}) is true for any consistent pair of vector and 
submultiplicative matrix norms. 

One appropriate set of such norms is the $l_1$ norm of a vector and the 
corresponding operator norm of a matrix. For a general $n$-vector $\mathbf{v}$ 
with components $v_i$ and a general  $n \times n$ matrix $\mathbfss{M}$ with 
entries $m_{ij}$, these norms are given respectively by 
\begin{equation}
\| \mathbf{v} \|_1 = \sum_{i=1}^n |v_i | \qquad \mathrm{and}
\qquad \| \mathbfss{M}\|_1 = \max_{1\leq j\leq n} \sum_{i=1}^n\left|m_{ij}\right|\,.
\label{eq:normdefs}
\end{equation}
The matrix norm $\| \mathbfss{M} \|_1$ corresponds to the maximum sum of the 
absolute values of the individual columns of $\mathbfss{M}$. 

Calculating matrix inverses are notoriously expensive.  Because an algorithm 
for estimating the $l_1$ norm of the inverse of a matrix is available, these norms 
have been used. The algorithm of \citet{1984hager} that estimates the $l_1$ norm 
of the inverse of a matrix without inverting the matrix first, was refined by 
\citet{1988higham}. This algorithm usually gives the exact value of $\| 
\mathbfss{A}^{-1}\|_1$ and, at worst, gives an order of magnitude estimate for 
the types of matrices considered here \citep{1988higham}. The values of $\| 
\mathbf{b}^{(i)}\|_1$ and $\| \mathbf{r}^{(i)}\|_1$ can be calculated directly.

\subsection{Effects on departure coefficients}

In this section an iterative procedure is applied to the linear 
system~(\ref{eq:saharatenl}) in order to show the effect of the stopping criterion 
given in equation~(\ref{eq:stop}) on the departure coefficients. The procedure is 
a derivative of the Gauss-Seidel method and is described in \citet{1971Brocklehurst}. 
The same method was used by \citetalias{1995Storey} and \citet{2017Salgado}. 

The results of the $n$-model are refined using this iterative 
scheme as follows. The departure coefficients $b_n$ from the $n$-model are used 
as the initial values so that $b_{nl}=b_n$ at the start of the algorithm. The rate 
equation~(\ref{eq:saharatenl}) is then solved to obtain $b_{nl}$ for decreasing 
values of $n$ starting with $n=n_\mathrm{c}$. The $b_{nl}$ values for a given $n$ 
are solved simultaneously as an $n \times n$ matrix by including all the terms 
that do not depend on these specific $b_{nl}$ values in the right-hand side of 
equation~(\ref{eq:nlsys}). Each newly calculated set of $b_{nl}$'s is used in 
subsequent calculations down to $n=2$ for Case A and $n=3$ for Case B. This 
constitutes one iteration and the process is repeated until the stopping 
criterion~(\ref{eq:stop}) is met.

\begin{figure}
	\includegraphics[width=\columnwidth]{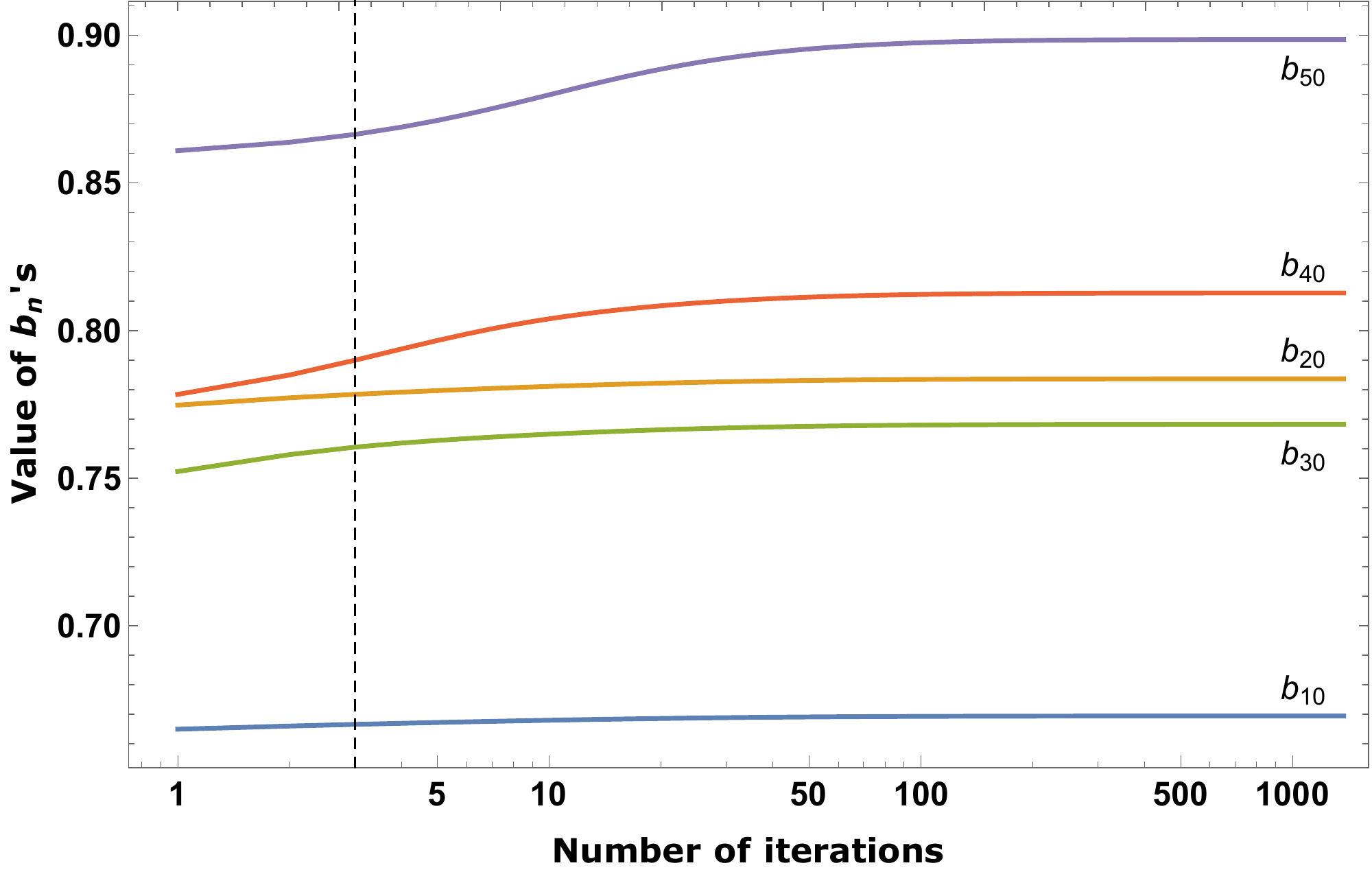}
		\caption{The evolution of five departure coefficients for 
$T_\mathrm{e}=10^4$\,K, $N_\mathrm{e}=10^4$\cmc\ Case B as the iterative procedure 
of the $nl$-model progresses. Convergence is reached after 1\,375
iterations. The dashed line at 3 iterations shows the first point 
where the values of the $b_{nl}$'s change by less than 1 per cent.}
\label{fig:iterations}
\end{figure}

Fig.~\ref{fig:iterations} shows the evolution of a subset of departure coefficients 
as the iterative procedure progresses for a gas at $T_\mathrm{e}=10^4$\,K with 
density $N_\mathrm{e}=10^4$\cmc\ for Case B with no external radiation field 
present. The values on the left-hand side of the graph show the values after one 
iteration and the values on the far right indicate the values after convergence 
is reached. For this case, the $b_{nl}$'s converged to four significant digits 
($\epsilon=5\times10^{-5}$) after $1\,375$ iterations. The dashed 
vertical line at 3 iterations indicates the first point in the process where the 
departure coefficients from two successive iterations change by less than 1 per 
cent. From the graph it is clear that a small change in the departure coefficients 
during the procedure is not sufficient to indicate convergence.

%%%%%%%%%%%%%%%%% RESULTS %%%%%%%%%%%%%%%%%%%%%

\section{Comparison to previous results}
\label{sec:comp}

\begin{figure}
	\includegraphics[width=\columnwidth]{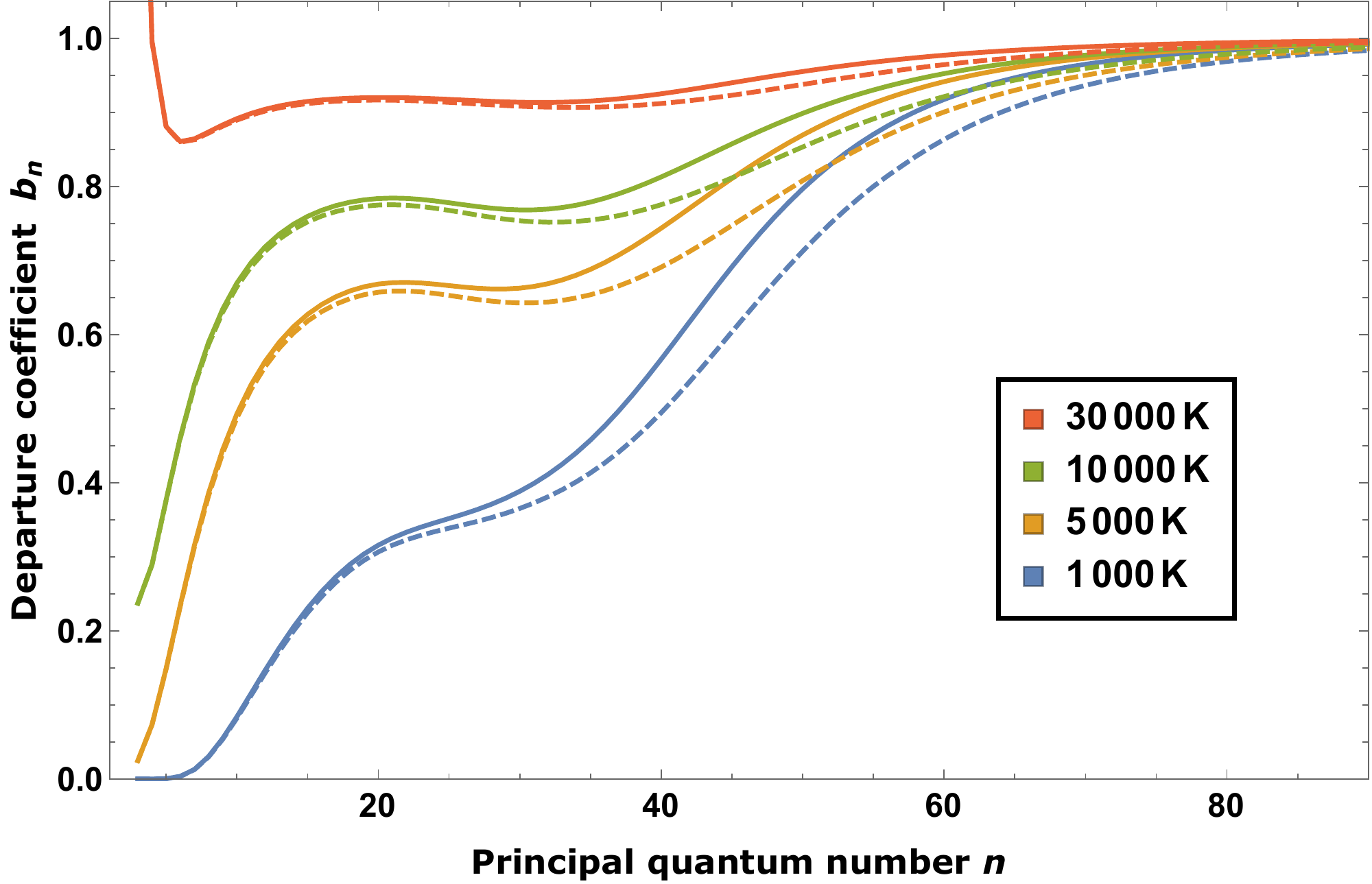}
		\caption{Departure coefficients $b_n$ for Case B for a variety of 
temperatures at electron density $N_\mathrm{e}=10^4$\cmc. Dashed lines represent 
the values obtained by \citetalias{1995Storey} and solid lines in the same colour 
show the current results for the same conditions. A colour version of this figure 
is available in the electronic version of the article.}
		\label{fig:tempdep}
\end{figure}

A comparison of the current results with those of \citetalias{1995Storey} for a 
range of temperatures at a fixed electron density of $N_\mathrm{e}=10^4$\cmc\ is 
shown in Fig.~\ref{fig:tempdep}. The largest discrepancy occurs at intermediate 
principal quantum numbers. This is unsurprising, since the behaviour of the $b_{nl}$'s 
at low and high $n$'s are not governed by the $nl$-model. At low energy levels, 
radiative processes dominate and the departure coefficients would be largely 
unaffected by the elastic collisions introduced in the $nl$-model. Therefore, at 
these levels the final $b_{nl}$'s will be very close to their initial values 
obtained from the $n$-model. At high energy levels, the collisional processes 
dominate completely and all $b_{nl}$'s will tend towards unity, reducing the 
difference between the two sets of calculations. 

\begin{figure}
	\includegraphics[width=\columnwidth]{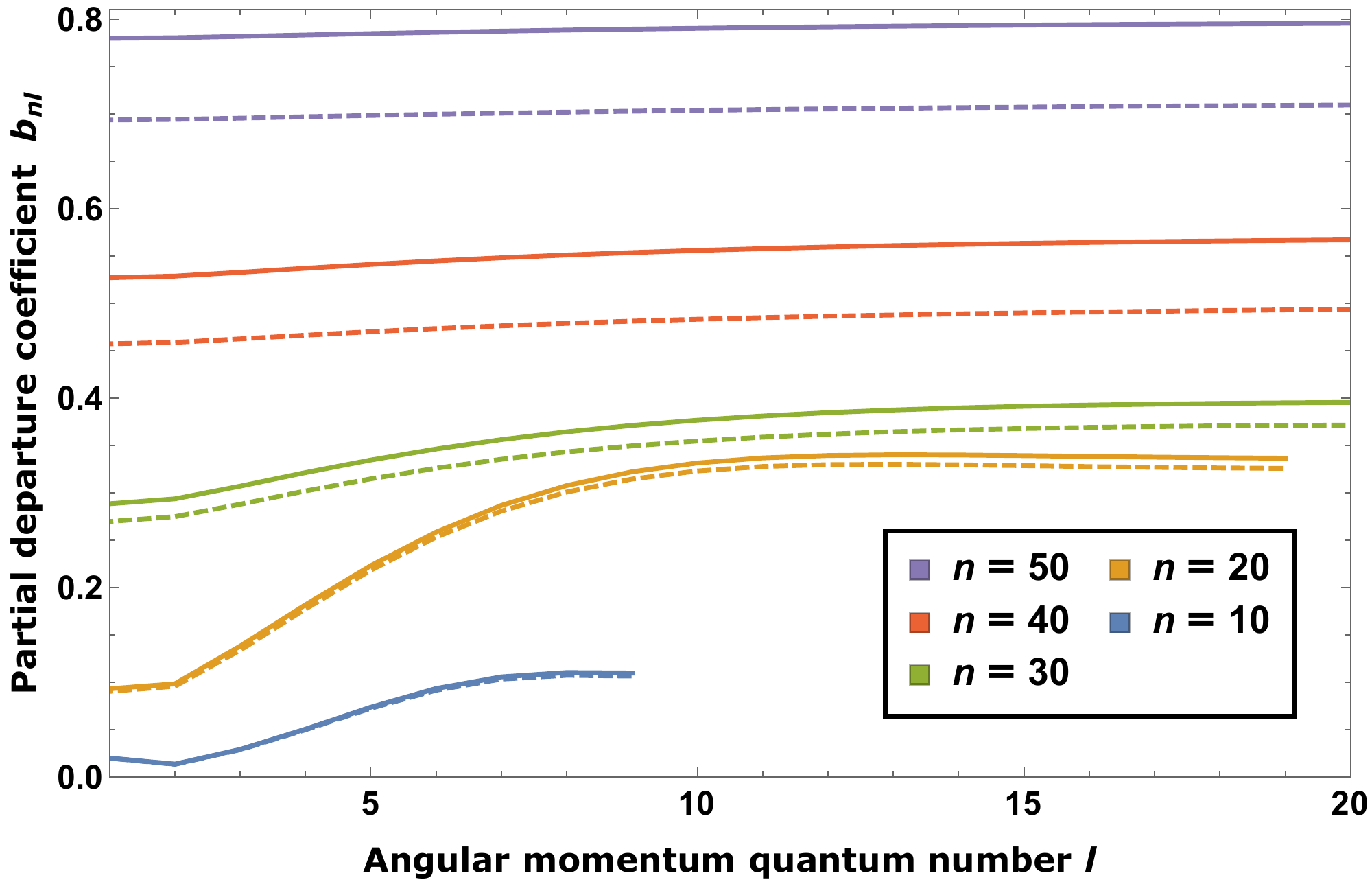}
		\caption{Partial departure coefficients $b_{nl}$ for a selected 
number of principal quantum numbers $n$ for $T_\mathrm{e}=10^3$\,K and 
$N_\mathrm{e}=10^4$\cmc, Case B. Solid lines represent the calculations of this 
work, dashed lines of the same colour show the results of \citetalias{1995Storey} 
for the same $n$. A colour version of this figure is available in the electronic 
version of the article.}
		\label{fig:ldep}
\end{figure}

Fig.~\ref{fig:ldep} shows the difference between the unsummed $b_{nl}$ values of 
\citetalias{1995Storey} and the current work. As expected, the differences are 
very small for low values of $n$, but become more pronounced as $n$ and $l$ 
increase. This effect is due to elastic collisions becoming the dominant process 
at intermediate and high $n$ levels.

\begin{figure}
	\includegraphics[width=\columnwidth]{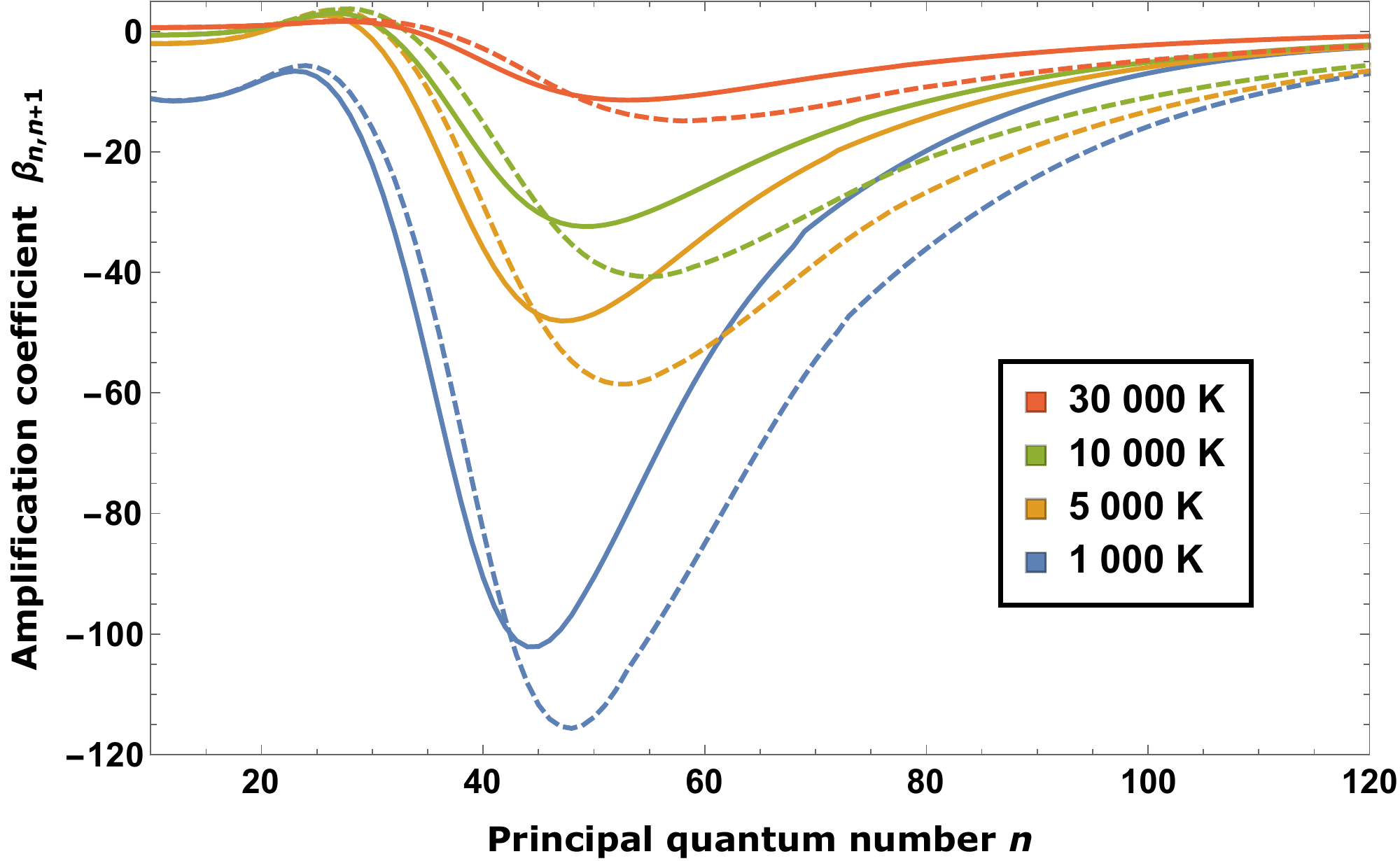}
		\caption{The amplification factor $\beta_{n,n+1}$ for H$n\alpha$ 
transitions for electron density $N_\mathrm{e}=10^4$\cmc\ and a range of temperatures 
under Case B conditions. Solid lines show the current results and dashed lines 
are those of \citetalias{1995Storey}. A colour version of this figure is available 
in the electronic version of the article.}
		\label{fig:betas}
\end{figure}

The results show that line enhancement by stimulated emission of H$n\alpha$ 
transitions for intermediate $n$ are less extreme than previously thought, with 
the point of maximum inversion occurring at a lower level. Fig.~\ref{fig:betas} 
illustrates the amplification factor as defined in equation~(\ref{eq:betadef}) 
for H$n\alpha$ transitions for the same parameters as Fig.~\ref{fig:tempdep}. 
The lines that fall in the optical range are largely unaffected, with the largest 
deviation from previous results  in the radio regime.

The discrepancy between our values and those of \citetalias{1995Storey} 
increases as the temperature decreases and the density increases; there is a 
stronger dependence on temperature than density. For $T_\mathrm{e} = 500$\,K and 
$N_\mathrm{e} = 10^6$\cmc\ the maximum relative difference is 22.5 per cent. We can 
match the values of \citetalias{1995Storey} within a per cent by terminating our 
procedure prematurely. Storey (private communication) has noted that the 
\citetalias{1995Storey} model converges extremely fast (within 10 iterations) and 
that the departure coefficients do not change significantly if either the number 
of iterations or $n_c$ is increased. 

At this stage we do not understand the reasons that the 
\citetalias{1995Storey} model converges so fast and to different values than ours. 
The differences in inelastic collisional rates between the models certainly plays 
a role, but does not account fully for the differences. We have performed many 
tests on our model, but cannot get it to converge as fast or to the same values 
as \citetalias{1995Storey}. Clearly, this is an issue that will need further 
investigation.

Obtaining a solution from thousands of iteration steps takes  a
significant amount of time. To 
speed up the solution, a direct solver using the 
PARDISO\footnote{\url{http://pardiso-project.org}} package \citep{pardiso-6.0b, 
pardiso-6.0a} was tested.  It is a sophisticated solver for systems of linear 
equations that exploits the sparsity of $\mathbfss{A}$ to solve the system in a 
very efficient manner. The direct solver is considerably faster at solving the 
set of linear equations than the iterative method.

The departure coefficients shown in the rest of the paper were obtained using the 
above direct solver package.  The $b_{nl}$'s obtained from the iterative method 
with the given stopping criterion match those using the direct solver. An error 
estimate of the quality of the results of the direct solver was obtained using
\begin{equation}
\epsilon \approx \| \mathbfss{A} \cdot \mathbf{b} - \mathbf{y} \| / \| \mathbf{b} \|\,.
\end{equation}
It was found that $\epsilon \leq 10^{-4}$ in all cases.

\label{sec:cont_fields}
%%%%%%%%%%%%%%%%% CONTINUUM FIELDS %%%%%%%%%%%%%%%%%%%%%

\section{Stimulating effects of continuum radiation}

In this section we examine the effects of the continuum radiation fields on the 
population structure of hydrogen in nebular environments. Specifically, their 
role in stimulating transitions between electronic states. 

The continuum radiation in the diffuse ISM consists of different components, each 
of which are examined to determine their effect on departure coefficients. The 
radiation fields considered here are the stellar radiation field, the free-free 
radiation field generated by the electrons in the gas, the cosmic microwave 
background radiation (CMBR) and emission from dust. 

\begin{figure}
	\includegraphics[width=\columnwidth]{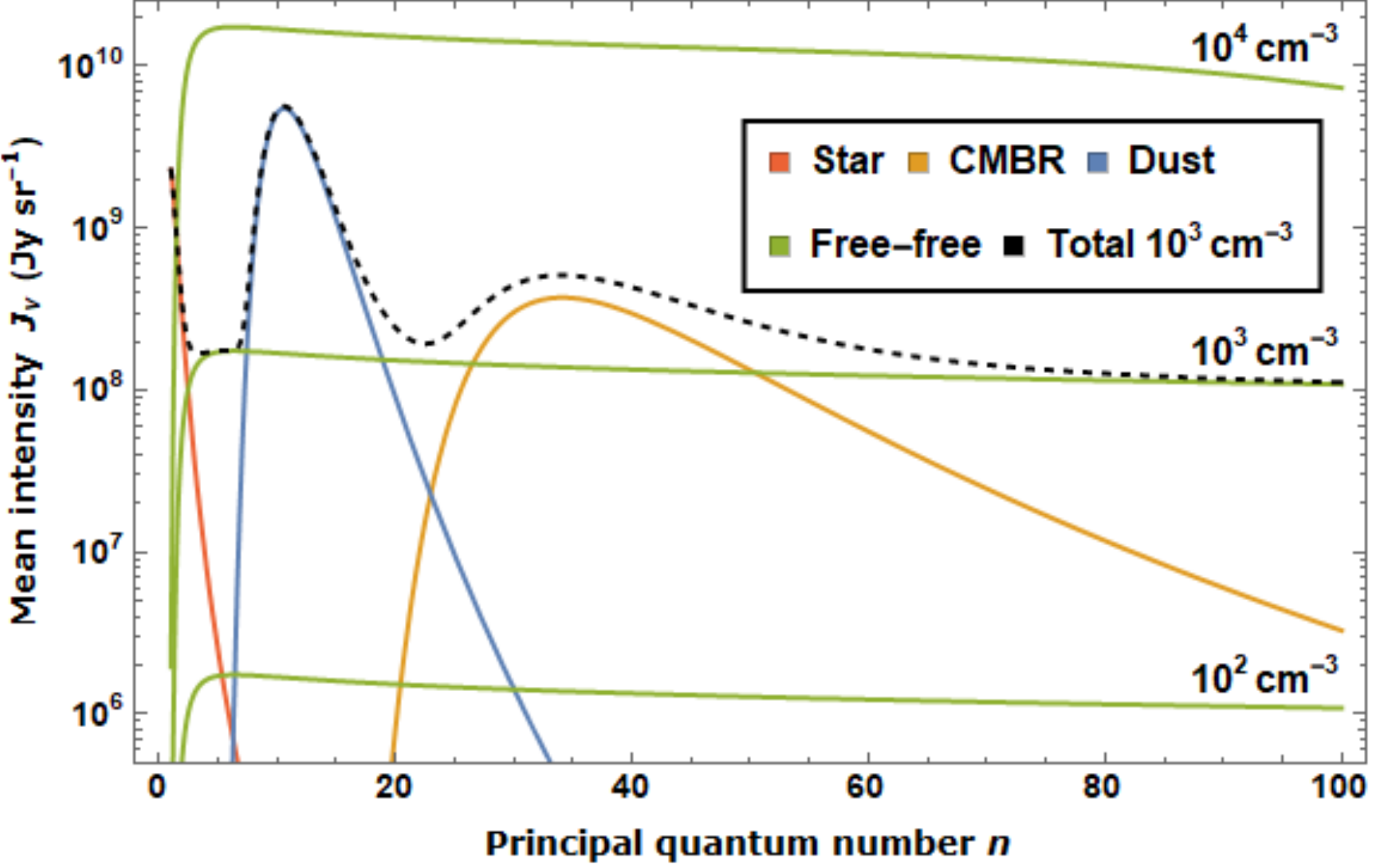}
		\caption{Spectrum of continuum radiation fields $J_{\nu}$ within 
a model ionised nebula at $T_\mathrm{e}=10^4$\,K. The free-free field is shown 
for three different densities. The horizontal axis shows the principal quantum 
number for H$n\alpha$ transitions, and the vertical axis shows the mean 
intensities of the various radiation fields.}
		\label{fig:cont_rad_fields}
\end{figure}

Fig.~\ref{fig:cont_rad_fields} shows the mean intensity $J_{\nu}$ of these fields 
as a function of the frequency associated with H$n\alpha$ transitions. The 
population inversion of the excited electrons is most pronounced for energy levels 
$30\lesssim n\lesssim80$ (refer to Fig.~\ref{fig:betas}), so that electrons in 
these levels are susceptible to being stimulated. 

\subsection{Stellar radiation field}

Photoionised nebulae require a nearby hot source of ultraviolet radiation to 
ionise atoms in the gas. In Fig.~\ref{fig:cont_rad_fields} the blackbody radiation 
field from a $T=50\,000$\,K source with dilution factor $W=10^{-12}$ is shown on 
the far left. As can be seen, the intensity of the ionising radiation from such 
a hot star drops off very quickly with decreasing frequency (increasing $n$ in 
the diagram) and is negligible at frequencies low enough to stimulate transitions
in hydrogen. An O or B star has to be a distance of $\lesssim10\,\mathrm{AU}$ from 
the gas to have any noticeable effect on the departure coefficients and, therefore, 
can be neglected in these calculations. 

\subsection{Cosmic microwave background radiation}

The CMBR has a low temperature ($T = 2.7$\,K) but is undiluted blackbody radiation. 
Coincidentally, the intensity of the radiation peaks around a frequency 
corresponding to H$n\alpha$ with $n \sim 40$ which is where the population 
inversion is strongest. 

For a density of $N_\mathrm{e} = 10^2$\cmc\ stimulated emissions due to the CMBR make up 
in excess of 10 per cent of the downward $\mathrm{H}n\alpha$ transitions for $40 
\lesssim n\lesssim60$. As the density increases, the effect of stimulated emission 
decreases because the influence of elastic collisions increase at these levels. 
The correction that the inclusion of the CMBR in the model provides to the summed 
$b_{nl}$'s is typically less than 1 per cent.

\subsection{Free-free radiation}

Continuous free-free radiation is produced by a plasma as charged free particles 
interact with each other without capture taking place. The free particles are 
assumed to be in thermodynamic equilibrium at a temperature $T_\mathrm{e}$. Disregarding 
background radiation, the specific intensity of the free-free radiation is given 
by 
\begin{equation}
J_{\nu}^{\mathrm{ff}} =B_{\nu}(T_\mathrm{e})\left(1-e^{-\tau_{\nu}^{\mathrm{ff}}}\right)\,,
\end{equation}
where $B_{\nu}$ is the Planck function for frequency, and $\tau_{\nu}^\mathrm{ff}$ is 
the optical depth of this radiation. 

Following \citet{2003Dickinson}, the appropriate optical depth for the free-free 
radiation is given by 
\begin{align}
\tau_{\nu}^{\mathrm{ff}} & = -3.014 \times 10^{-2}\, T_\mathrm{e}^{-3/2} \left( 
\frac{10^{9}}{\nu} \right)^2 \nonumber\\
& \times \left[ \ln\left(4.955 \times 10^{-2} \nu^{-1}\right) + 1.5\ln\left(T_\mathrm{e} 
\right) \right](\mathrm{EM})\,,
\end{align}
where EM is the emission measure in cm$^{-6}$\,pc. For a homogeneous gas, EM = 
$N_\mathrm{e}^2$\,cm$^{-6}$\,pc.

\begin{figure}
	\includegraphics[width=\columnwidth]{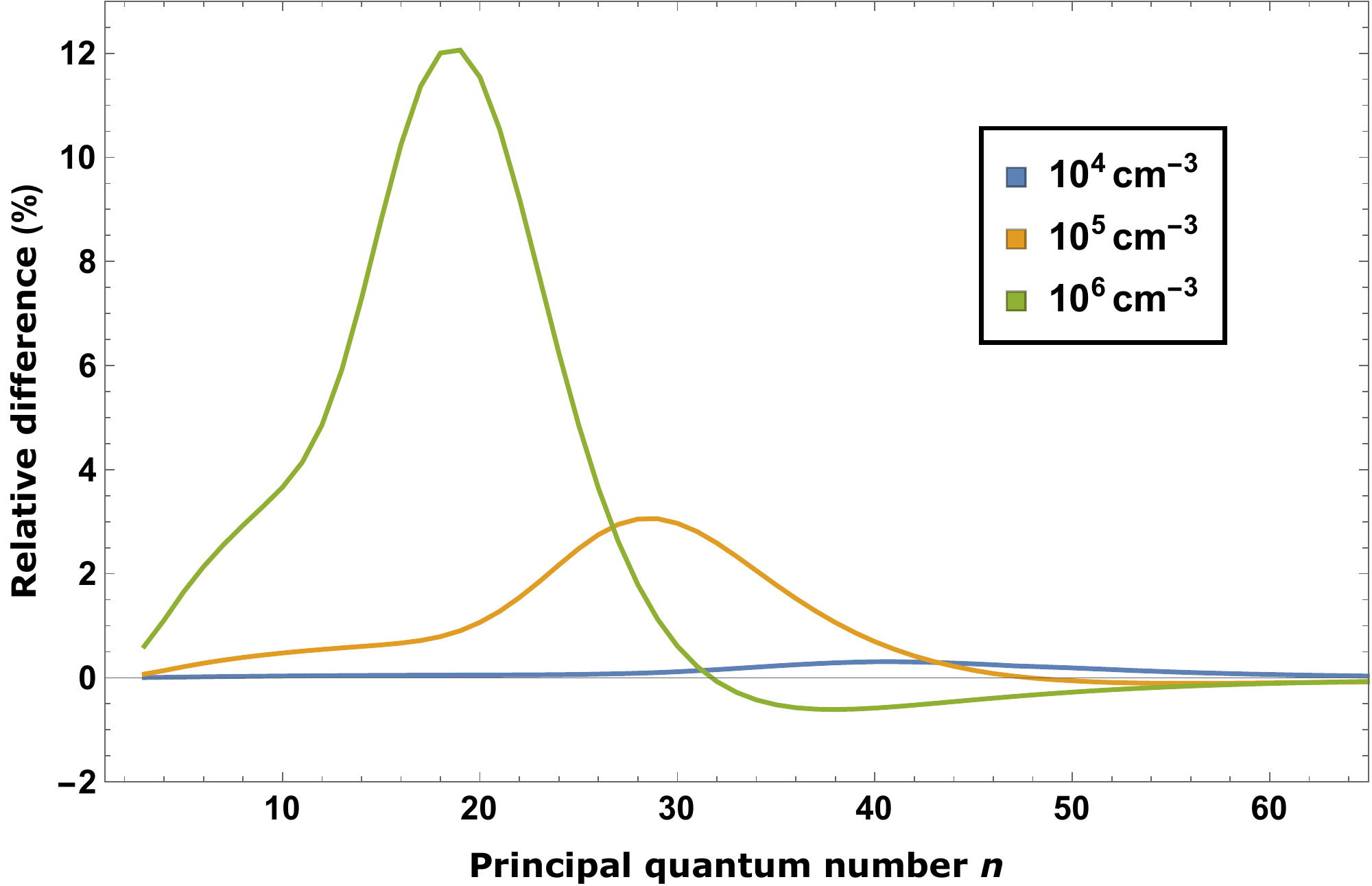}
		\caption{The effects of the free-free radiation field on the 
population structure of hydrogen for $T_\mathrm{e}=10^4$\,K, and a range of 
densities in Case B\@. The relative difference between the departure coefficients 
calculated when no radiation field is present and when the free-free field is 
included in the calculations are shown for each principal quantum number. }
		\label{fig:effect_ff}
\end{figure}

The intensity of the free-free radiation within a nebula is strongly dependent on 
the electron density $N_\mathrm{e}$. The effect of this radiation on the population 
structure of hydrogen is negligible for low densities, but can become significant 
for $N_\mathrm{e} > 10^4$\cmc. For example, the departure coefficients of a gas 
with $N_\mathrm{e} = 10^6$\cmc\ and $T_\mathrm{e} = 10^4$\,K will be affected by 
up to 12 per cent around $n=20$ by the inclusion of the free-free radiation field 
due to stimulated processes, as illustrated in Fig.~\ref{fig:effect_ff}. The 
free-free radiation affects departure coefficients for principal quantum numbers 
in the range $10\lesssim n\lesssim60$.

\subsection{Dust}

Dust grains form an important component of the ISM and have been found to be 
intermixed with the ionised media of \HII\ regions and planetary nebulae 
\citep{1993Barlow,1997Kingdon}. Emission from dust grains can dominate the spectrum 
from \HII\ regions and PNe at long wavelengths, outshining the free-free specific 
intensity by orders of magnitudes. 

To model the emission from dust within the cloud, a modified blackbody spectrum 
of the form
\begin{equation}
J_{\nu}^\mathrm{d} = \tau_{\nu}^\mathrm{d} B_{\nu}(T_\mathrm{d})\,,
\end{equation}
was used, where $\tau_{\nu}^\mathrm{d}$ is an optical depth and $T_\mathrm{d}$ is 
the dust temperature \citep{2014Planck}. The angle-averaged optical depth 
\begin{equation}
\tau_{\lambda}^\mathrm{d} =1.5 \times 10^{-3} \left(\frac{100\mathrm{\mu m}}{\lambda}
\right)^{1.7}
\end{equation}
from \citet{2011Draine} was used.

The effect of the dust radiation on the $b_n$ values is very limited as the 
populations are only weakly inverted at the frequencies where this radiation can 
stimulate transitions. In Fig.~\ref{fig:cont_rad_fields} a dust temperature of 
$T_\mathrm{d} = 50$\,K \citep{2003Dupac} is shown, but the result is independent 
of $T_\mathrm{d}$ since this has little effect on the frequency range of the field. 

In this work we have only considered the effect of the radiation 
from dust on the level populations. \citet{1992Hummer} have shown that absorption 
by dust can affect the Case B recombination spectrum, possibly with a greater 
effect on the departure coefficients than emission.

%%%%%%%%%%%%%%%%%% CONCLUSIONS %%%%%%%%%%%%%%%%%%%%%

\section{Description of tables}
The programme described here calculates departure coefficients $b_{nl}$ for level 
$nl$ in hydrogen. From this, theoretical spectral line intensities can be 
calculated. The formulae required to calculate values that can be compared with 
observed lines are presented below, and then the entries in the tables are 
explained.

\begin{table*}
\ttfamily
\resizebox{\textwidth}{!}{%
\begin{tabular}{rrrrrrrrrrrrrrrrrrrr}
\multicolumn{20}{l}{NE = 10000 TE = 10000 CASE B NC = 210 NO RAD}\\
\multicolumn{20}{l}{-------------------------------------------------------------------}\\
\multicolumn{20}{l}{BN'S}\\
3 & 2.374e-01 & 4 & 2.895e-01 & 5 & 3.769e-01 & 6 & 4.616e-01 & 7 & 5.328e-01 & 8 & 5.897e-01 & 9 & 6.344e-01 & 10 & 6.694e-01 & 11 & 6.968e-01 & 12 & 7.184e-01\\
13 & 7.355e-01 & 14 & 7.489e-01 & 15 & 7.595e-01 & 16 & 7.678e-01 & 17 & 7.741e-01 & 18 & 7.787e-01 & 19 & 7.818e-01 & 20 & 7.836e-01 & 21 & 7.842e-01 & 22 & 7.838e-01\\
$\boldsymbol{\vdots}$ &  &  &  &  &  &  &  &  &  &  &  &  &  &  &  &  &  &  & \\
\multicolumn{14}{l}{BNL'S} &  &  &  &  &  & \\
\multicolumn{14}{l}{n = 3} &  &  &  &  &  & \\
0 & 1.022e+00 & 1 & 2.336e-01 & 2 & 8.271e-02 &  &  &  &  &  &  &  &  &  &  &  &  &  & \\
\multicolumn{14}{l}{n = 4} &  &  &  &  &  & \\
0 & 1.132e+00 & 1 & 3.300e-01 & 2 & 1.446e-01 & 3 & 2.553e-01 &  &  &  &  &  &  &  &  &  &  &  & \\
\multicolumn{14}{l}{n = 5} &  &  &  &  &  & \\
0 & 1.206e+00 & 1 & 3.973e-01 & 2 & 1.891e-01 & 3 & 3.235e-01 & 4 & 4.239e-01 &  &  &  &  &  &  &  &  &  & \\
$\boldsymbol{\vdots}$  &  &  &  &  &  &  &  &  &  &  &  &  &  &  &  &  &  &  & \\
\multicolumn{20}{l}{JNM'S}\\
\multicolumn{20}{l}{n = 4}\\
1 & 1.217e-25 & 2 & 9.879e-27 & 3 & 3.289e-27 &  &  &  &  &  &  &  &  &  &  &  &  &  & \\
\multicolumn{20}{l}{n = 5}\\
1 & 5.304e-26 & 2 & 4.633e-27 & 3 & 1.602e-27 & 4 & 7.710e-28 &  &  &  &  &  &  &  &  &  &  &  & \\
\multicolumn{20}{l}{n = 6}\\
1 & 2.830e-26 & 2 & 2.563e-27 & 3 & 8.903e-28 & 4 & 4.426e-28 & 5 & 2.432e-28 &  &  &  &  &  &  &  &  &  & \\
$\boldsymbol{\vdots}$  &  &  &  &  &  &  &  &  &  &  &  &  &  &  &  &  &  &  & \\
\multicolumn{20}{l}{KMN'S}\\
\multicolumn{20}{l}{n = 4}\\
3 & 3.271e-23 &  &  &  &  &  &  &  &  &  &  &  &  &  &  &  &  &  & \\
\multicolumn{20}{l}{n = 5}\\
3 & 1.003e-23 & 4 & -8.418e-25 &  &  &  &  &  &  &  &  &  &  &  &  &  &  &  & \\
\multicolumn{20}{l}{n = 6}\\
3 & 4.463e-24 & 4 & 2.892e-24 & 5 & -1.518e-23 &  &  &  &  &  &  &  &  &  &  &  &  &  & \\
$\boldsymbol{\vdots}$  &  &  &  &  &  &  &  &  &  &  &  &  &  &  &  &  &  &  & \\
\end{tabular}}
\caption{\label{tab:sample_table}An extract of one of the output tables for the 
conditions $T_\mathrm{e}=10^4$\,K and $N_\mathrm{e}=10^4$\cmc, Case B, with no external radiation 
present. The values of the departure coefficients ($b_n$), the partial departure 
coefficients ($b_{nl}$), the emission coefficients ($j_{nn'}$) and the absorption 
coefficients ($\kappa_{n'n}$) are tabulated for each case.} 
\end{table*}

The emission coefficient $j_{nn'}$ for line radiation is given by 
\begin{equation}
j_{nn'}=\frac{h\nu}{4\pi}\sum_{l=0}^{n-1}\sum_{l'=l\pm1}b_{nl}N_{nl}^*A_{nl,n'l'}
\label{eq:emcoef}
\end{equation}
and the absorption coefficient by
\begin{equation}
\kappa_{n'n} = \frac{h\nu}{4\pi} \sum_{l=0}^{n-1}\sum_{l'=l\pm1}\left[ b_{nl} 
N_{nl}^*B_{n'l',nl} \left( 1-\frac{b_{nl}}{b_{n'l'}} e^{-h\nu/k_\mathrm{B}T_\mathrm{e}} \right) 
\right]\,.
\label{eq:abscoef}
\end{equation}

The tables containing our results are available online in machine-readable ASCII 
format and can be downloaded from the article webpage. The file names are a 
concatenation of the significands and exponents of the electron density and 
temperature, the case (A or B) and any ambient radiation that has been included 
in the model, all separated by an underscore. Free-free radiation is indicated by
`FF', the CMBR by `CMB', and no radiation field is designated with the number `0'. 
For example, the file named `13\_14\_B\_0.dat' contains the results for Case B 
with $N_\mathrm{e}=10^4$\cmc\ and $T_\mathrm{e}=10^4$\,K with no ambient 
radiation present. The header in each file contains the same data as in the file name, 
as well as the value of $n_\mathrm{c}$.

Each file contains the $b_n$ values calculated using equation~(\ref{eq:bnsum}) 
for each value of $n$ from $n=2$ for Case A and $n=3$ for Case B to $n=500$. This 
is followed by the partial departure coefficients $b_{nl}$ with the appropriate 
values of $l$ tabulated after each value of $n$ up to $n=100$. The emission 
coefficients $j_{nn'}$ and absorption coefficients $\kappa_{n'n}$ as defined by 
equations~(\ref{eq:emcoef}) and (\ref{eq:abscoef}), respectively, are given next. 
The coefficients are tabulated next to the lower level $n'$ for ascending values 
of the upper level $n$. An extract of one of the data files is shown in table~\ref{tab:sample_table}.

%%%%%%%%%%%%%%%%%% CONCLUSIONS %%%%%%%%%%%%%%%%%%%%%

\section{Conclusions}
\label{sec:conclusions}

Updated calculations of departure coefficients for hydrogen atoms under nebular 
conditions are presented. The elastic collision rates of \citet{2016Guzman} have 
been used, which differ from values used in previous models. We have also included 
stimulated and absorption processes in our equations of statistical equilibrium. 
The model used to do the calculations is similar to that of \citetalias{1995Storey}.  

A stopping criterion has been derived and employed to determine when to terminate 
the iterative procedure, which ensures that the $b_{nl}$ values have converged to 
a predefined accuracy. This requires that many more iterative steps are required 
before a solution is reached than have been used in previous works. In practice, 
we found that it is more time efficient to use a direct solver rather than the 
iterative method to achieve an acceptable accuracy. 

We investigated the effects of stimulated emission due to various components of 
the continuum field within a nebula on the population structure of hydrogen. Even 
though the stellar radiation field and emission from dust dominate the continuum 
spectrum at certain frequencies, the effects on the population structure of both 
fields are negligible. The free-free radiation field has the largest influence on 
the departure coefficients, increasing as the density increases. The CMBR typically 
has an effect of less than 1 per cent on the departure coefficients. 

Our results give departure coefficients that are consistently larger (closer to 
LTE) than those of \citetalias{1995Storey}. The current results produce 
amplification factors $\beta_{n,n+1}$ that are smaller than in previous 
calculations, producing less extreme population inversions. The value of $n_\mathrm{c}$, 
the level at which populations become statistically distributed, has been 
determined empirically. The value, which depends on the electron density $N_\mathrm{e}$, 
is much larger than used in other models.

Results for He and metal atoms and ions will be considered in a separate publication.

\section*{Acknowledgements}
We thank the reviewer, Prof P. J. Storey, for his insightful comments on our paper.
AP thanks Unisa for the funding she has received from the Academic Qualification 
Incentive Programme.

%%%%%%%%%%%%%%%%%%%%%%%%%%%%%%%%%%%%%%%%%%%%%%%%%%



\bibliographystyle{mnras}
\bibliography{mnras_art} % if your bibtex file is called example.bib

%%%%%%%%%%%%%%%%%%%%%%%%%%%%%%%%%%%%%%%%%%%%%%%%%%

%%%%%%%%%%%%%%%%% APPENDICES %%%%%%%%%%%%%%%%%%%%%

\appendix

\section{Calculational details}
\label{sec:calcs}
The various atomic rates used in our calculations are based on standard algorithms 
from the literature which, in many cases, have been used by numerous authors over 
the years. There are possibly slight differences in formulae used for large values 
of $n$ and the values at which these approximations are used, and therefore, for 
completeness, we present in this section how we calculated the various rates.  
Because there is as yet no compelling evidence to indicate that particles in 
astronomical plasmas obey anything other than regular thermodynamic velocity 
distributions, all our velocity distribution functions are Maxwellian.  

Although the $n$-method models are run before the $nl$-models, we usually 
calculate the $nl$-model atomic rates first, and then use these with the 
appropriate weighting to determine the $n$-model value. The conversion formulae 
we used are presented at the end of each section. 

The symbols used are: $a_0$ is the Bohr radius, $c$ is the speed of light in a 
vacuum, $e$ is the elementary charge, $g_{nl}$ is the statistical weight of level 
$nl$, $h$ is Planck's constant, $k_\mathrm{B}$ is Boltzmann's constant, $m_\mathrm{e}$ 
is the electron mass, $R_\mathrm{H}$ is the Rydberg constant for hydrogen,  
$T_\mathrm{e}$ is the kinetic temperature of the free electron gas, $Z$ is the 
atomic charge, $\alpha$ is the fine structure constant, and $\nu$ is the frequency 
of the emitted or absorbed photon.

\subsection{Radiative processes}
Because electric dipole transitions are orders of magnitude larger than forbidden 
transitions, only transitions that satisfy $l\rightarrow l \pm1$ are included in 
our models. 
\subsubsection{Bound-bound transitions}

The rate of a spontaneous dipole transition from level $nl$ to a lower energy 
level $n'l'$ is calculated using \citep{1971Brocklehurst}

\begin{equation}
\label{A1}
A_{nl,\,n'l'} = \frac{64 \pi^4 \nu^3}{3Z^2 hc^3} \frac{\max(l,\,l')}{2l+1} e^2 
a_0^2 \left[ \tau_{nl}^{n'l'} \right]^2
\end{equation}
in cgs units, where $\tau_{nl}^{n'l'}$ is the bound-bound radial dipole matrix 
element of the transition. An explicit formula for the bound-bound dipole matrix 
element of a transition between $nl$ and $n'\,l-1$ for hydrogen is given by 
\citet{1929Gordon}. The matrix element $\tau_{nl}^{n'\,l+1}$ for a transition 
$nl\rightarrow n'\,l+1$ is obtained by making the substitutions $n\rightarrow n'$ 
and $l\rightarrow l+1$ in the above formula.

Direct calculation of the matrix elements $\tau_{nl}^{n'l'}$ is impractical for 
large principal quantum numbers due to overflow errors in the calculation of the 
hypergeometric functions. An iterative scheme described by \citet{1979vanRegemorter} 
is used to calculate $\tau_{nl}^{n'l'}$ for large $n$. The results were compared 
to the tables of \citet{1957Green}, which are almost identical to the relativistic 
transition probabilities of \citet{2009Wiese}, and show good agreement. 

The Einstein A-values for the $n$-model, $A_{nn'}$, are obtained from (\ref{A1}) 
using 
\begin{equation}
A_{nn'}=\frac{1}{n^2}\sum_l \sum_{l'=l\pm1}\left(2l+1\right)A_{nl,\,n'l'}\,.
\end{equation}

For levels with $n > 500$, the approximation due to \citet{1937Menzel} for the 
A-value is used, which is given by
\begin{equation}
A_{nn'} = \frac{16 \alpha^4 c}{3\pi \sqrt{3} a_0} \frac{g_{nn'}}{n^3 n'(n^2 - n'^2)}
\label{eq:avalaprox}
\end{equation}
where $g_{nn'}$ is the bound-bound Gaunt factor that is approximated by 
\begin{equation}
g_{nn'} = 1 - \frac{0.1728 \left(1+n'^2/n^2 \right)}{\left(1-n'^2/n^2\right)^{2/3}
\left(n'\right)^{2/3}}\,.
\end{equation}

For high $n$-levels, the approximation (\ref{eq:avalaprox}) generally performs 
well. Because collisional processes will always dominate over radiative processes 
at these levels (for the conditions considered here), inaccuracies introduced by 
this approximation are not important. 

The rate coefficients for  stimulated emission $B_{nl,\,n'l'}$ and absorption 
$B_{n'l',\,nl}$ are calculated using the Einstein relations \citep{1916Einstein} 
given by
\begin{equation}
B_{nl,\,n'l'} = \frac{g_{n'l'}}{g_{nl}} B_{n'l',\,nl} = \frac{c^2}{2h \nu^3}
A_{nl,\,n'l'}\,.
\label{eq:einstein stim}
\end{equation}

\subsubsection{Bound-free transitions}
\paragraph{Radiative recombination}

The radiative recombination rate coefficient $\alpha_{nl}^\mathrm{r}$ to the level 
$nl$ is defined by 
\begin{equation}
N_\mathrm{e} N_\mathrm{p} \alpha_{nl}^\mathrm{r} = N_\mathrm{e} N_\mathrm{p} \int 
\sigma_{nl}^r(v)\,vf(v)\,dv \label{eq:radrecomdef}
\end{equation}
where $\sigma_{nl}^\mathrm{r}(v)$ is the radiative recombination cross-section 
and $f(v)$ is the speed distribution of the free electrons. 

An expression for $\alpha_{nl}^\mathrm{r}$ that uses a Maxwellian distribution 
with a temperature $T_\mathrm{e}$ for the free electron velocities is given by 
\citet{1965Burgess} as
\begin{equation}
\alpha_{nl}^\mathrm{r} = \left( \frac{2\pi^{1/2}\alpha^4 a_0^2 c}{3} \right) 
\frac{2y^{1/2}}{n^2} \sum_{l'=l\pm1}I\!\left(n,\,l,\,l',\,T_\mathrm{e}\right)
\end{equation}
where
\begin{align}
I\!\left(n,\,l,\,l',\,T_\mathrm{e}\right) & = \mathrm{max}\!\left(l,\,l'\right) 
\,y \nonumber \\ 
& \times \intop_0^{\infty} \left(1+n^2\kappa^2\right)^2 \Theta\!\left(n,\,l;\,
\kappa,\,l'\right) \mathrm{e}^{-y\kappa^2}\;d\!\left(\kappa^2\right)
\label{eq:radint}
\end{align}
and
\begin{equation}
y=\frac{R_\mathrm{H} hc}{k_\mathrm{B} T_\mathrm{e}},\qquad\Theta(n,\,l;\,\kappa,\,l') 
= \left(1+n^2\kappa^2 \right) \left|\tau_{nl}^{\kappa l'}\right|^2\,.
\end{equation}

The parameter $\kappa$ is the energy of the free electron expressed in terms of 
the ionisation energy of hydrogen so that the energy difference between the bound 
and free states is given by
\begin{equation}
h\nu = \left(1+n^2 \kappa^2\right) \chi_n
\label{eq:defk}
\end{equation}
with $\chi_n = \chi_1/n^2$ the ionisation energy of level $n$.

The radial dipole matrix elements for transitions between a bound and a free state 
$\tau_{nl}^{\kappa l'}$ are computed using the recurrence relations given by 
\citet{1965Burgess}. These equations satisfy the exact matrix elements for 
hydrogenic atoms. \citet{1965Burgess} observes that severe scaling problems can 
occur with this method for large $n$. This was overcome by doing calculations with 
the natural logarithm of the recurrence relations. The natural logarithm reduces 
the size of the numbers stored in the computer memory at intermediate steps, 
thereby preventing overflow errors. 

The integration in (\ref{eq:radint}) is performed with a Gaussian integration 
technique. The integrands vary very rapidly when $\kappa$ is small, and much slower 
for large $\kappa$. Following \citet{1965Burgess}, a number of five-point Gaussian 
integrations are made, starting with an interval size of $h=10^{-4}n^{-1}$. The 
interval size was doubled after every five-point integration and the procedure is 
terminated when the sum of the integrals is accurate up to six significant digits. 

The total radiative recombination coefficient into an energy level $n$ is given 
by the sum of the partial coefficients over the angular momentum states so that 
\begin{equation}
\alpha_n^\mathrm{r} = \sum_{l=0}^{n-1}\alpha_{nl}^\mathrm{r}\,.
\end{equation}

\paragraph{Photoionisation}

The rate coefficient for photoionisation $\alpha_{nl}^\mathrm{p}(\nu)$ for the 
level $nl$ is given by 
\begin{equation}
N_{nl}\alpha_{nl}^\mathrm{p} = N_{nl}\int_{\chi_{nl}/h}^{\infty}a_{nl}^\mathrm{p}
(\nu)\frac{4\pi J_{\nu}}{h\nu}\,d\nu\,.\label{eq:phiondef}
\end{equation}
where $J_{\nu}$ is the mean intensity of the ambient radiation field and 
$a_{nl}^\mathrm{p}(\nu)$ is the photoionisation cross section from level $nl$ for 
a photon with frequency $\nu$.

The photoionisation cross-section to level $nl$ for a hydrogenic atom is calculated 
using the formula of \citet{1965Burgess}, which is given by
\begin{equation}
a_{nl}^\mathrm{p}\!\left(\kappa^2\right) = \left(\frac{4\pi\alpha a_0^2}{3}\right) n^2
\sum_{l'=l\pm1}\frac{\max(l,\,l')}{2l+1}\,\Theta\!\left(n,\,l;\,\kappa,\,l'\right)\,.
\label{eq:phioncs}
\end{equation}

Putting equation~(\ref{eq:phioncs}) into (\ref{eq:phiondef}) and using equation~(\ref{eq:defk}),
the photoionisation coefficient $\alpha_{nl}^\mathrm{p}$ can be written as
\begin{align}
\alpha_{nl}^\mathrm{p} &= \left(\frac{4\pi R_\mathrm{H} c\alpha a_0^2}{3}\right) \frac{n^2}{2l+1} 
\int_0^{\infty} \left(\frac{4\pi J_{\nu}}{h\nu}\right)_{\kappa^2} \nonumber\\
& \times \sum_{l'=l\pm1}\max(l,\,l')\,\Theta\!\left(n,\,l;\,\kappa,\,l'\right)\,
d\left(\kappa^2\right)\,.
\label{eq:phiocoef}
\end{align}
The subscript $\kappa^2$ indicates that the quantity inside the brackets should 
be written in terms of $\kappa^2$ before integration. The integral in 
(\ref{eq:phiocoef}) is handled in the same manner as for radiative recombination.

The total photoionisation coefficient for hydrogen from level $n$ is given by
\begin{equation}
\alpha_n^\mathrm{p} = \sum_{l=0}^{n-1}\frac{2l+1}{n^2}\alpha_{nl}^\mathrm{p}\,.
\end{equation}

\paragraph{Stimulated recombination}

The cross-section for stimulated recombination $\sigma_{nl}^\mathrm{s}(v)$ of an electron 
with a speed $v$ into level $nl$ is related to the photoionisation cross-section 
$a_{nl}^\mathrm{p}$ (see equation~(\ref{eq:phioncs})) by
\begin{equation}
\sigma_{nl}^s = \frac{g_{nl}}{8\pi g_i} \frac{h^2}{m_\mathrm{e}^2v^2} a_{nl}^\mathrm{p}\,.
\label{eq:stimcsE-M}
\end{equation}
The rate of stimulated recombination in the presence of a radiation field $J_{\nu}$ 
with a free electron velocity distribution $f(v)$ is given by
\begin{equation}
\alpha_{nl}^\mathrm{s} = \int_{\chi_{nl}/h}^{\infty} \frac{4\pi J_{\nu}}{h\nu} 
\sigma_{nl}^\mathrm{s}(v)\,f(v) \frac{h}{m_\mathrm{e}}\,d\nu\,.\label{eq:stimrecombdef}
\end{equation}

Putting equation~(\ref{eq:phioncs}) into (\ref{eq:stimcsE-M}) gives the stimulated 
recombination cross-section, and in turn putting the result into 
equation~(\ref{eq:stimrecombdef}) gives the stimulated recombination coefficient as
\begin{align}
\alpha_{nl}^\mathrm{s} & =  \left( \frac{4\pi R_\mathrm{H} c\alpha a_0^2}{3} \right) \left( \frac{m_\mathrm{e}}
{2\pi k_BT_\mathrm{e}} \right)^{3/2} \left( \frac{h}{m_\mathrm{e}} \right)^3n^2\nonumber\\
 &  \times \int_0^{\infty} \left(\frac{4\pi J_{\nu}}{h\nu}\right)_{\kappa^2} 
\sum_{l'=l\pm1}\max(l,\,l')\,\Theta\!\left(n,\,l;\,\kappa,\,l'\right)
e^{-y\kappa^2}d(\kappa^2)
\end{align}
The integration is evaluated numerically using a Gaussian quadrature scheme as 
described above for radiative recombination.

The total stimulated recombination coefficient to an energy level $n$ of hydrogen 
is given by
\begin{equation}
\alpha_n^\mathrm{s} = \sum_{l=0}^{n-1} \alpha_{nl}^\mathrm{s}\,.
\end{equation}

%%%%%%%%%%%%%%%%%%%%%%%%%%%

\subsection{Collisional transitions}

\subsubsection{Inelastic collisions}

The semi-empirical formulae of \citet{1980VriensSmeets} are used to calculate the 
rates for collisional de-excitation $C_{n,n'}$ between the bound states $n$ and 
$n'$. These values are valid over a wider range of temperatures than the values 
obtained from the formulae of \citet{1976Gee} which were used by 
\citetalias{1995Storey}. \citet{1980VriensSmeets} claim that their results agree 
within 5 to 20 per cent with those of \citet{1976Gee} in the regimes where both 
calculations are valid. 

\citetalias{1995Storey} resolved the collision rates between angular momentum 
states using
\begin{equation}
C_{nl,\,n'l'}  =  \frac{A_{nl,\,n'l'}}{A_{nn'}}C_{nn'}\label{eq:colanlml}
\end{equation}
for $\Delta l=\pm1$, which is based on the Coulomb-Bethe approximation. 
\citetalias{1995Storey} used this approximation for transitions with $n' = n\pm1$ 
and 
\begin{equation} 
C_{nl,\,n'l'}  =  C_{nn'} \quad \mbox{for}\quad n' \neq n \pm 1 \,. 
\end{equation}
In our calculations the approximation in equation~(\ref{eq:colanlml}) is used 
throughout; detailed balance is used to calculate the rates of the inverse process. 
We included interactions between all levels with $n \neq n'$.
Because elastic collision rates are higher than inelastic collision rates for all 
$n$ for which collisions are important, $l$-levels are populated predominantly by 
angular momentum-changing collisions, and, hence, the exact dependence of the 
inelastic collision rates on angular momentum is not that important.

We use the rates of collisional ionisation $C_{n,i}$ from the formulae of 
\citet{1980VriensSmeets}. The assumption $C_{nl,i} = C_{n,i}$ is used in the 
$nl$-model, since this process is never significant at the conditions considered 
here. The three-body recombination rates $C_{i,n}$ and $C_{i,nl}$ are obtained 
from detailed balance considerations. \citet{1980VriensSmeets} found that their 
formulae for the bound-free collisional rates agreed with experimental data to 
within 10 to 30 per cent.

\subsubsection{Elastic collisions}
\label{sec:elascol}

\citet{1971Brocklehurst} used an iterative scheme to calculate the partial rate 
coefficients $C_{nl,nl\pm1}$ based on the work of \citet{1964PengellySeaton}. 
\citet{1987HummerStorey} pointed out that this method caused oscillatory behaviour 
in the rate coefficients as a function of $l$. They avoided this by normalizing 
the collision cross-sections with the oscillator strengths of the transitions.

In this work, the formalism of \citet{1964PengellySeaton} is followed based on 
the recommendation of \citet{2015StoreySochi} and \citet{2016Guzman}. The 
modification suggested by \citet{2016Guzman} to get the partial rates directly 
is utilized.  The relevant equations in cgs units are 
\begin{align}
C_{nl,nl} & = 9.93\times10^{-6} \sqrt{\frac{\mu}{m}} \frac{D_{nl,nl'}}{\sqrt{T_\mathrm{e}}} 
N_\mathrm{e} \nonumber\\
 & \times \left[ 11.54 + \log_{10} \left( \frac{mT_\mathrm{e}}{\mu D_{nl,nl}} \right) + 
2\log_{10} \left(R_\mathrm{c} \right) \right] \label{eq:momcolrec2}
\end{align}
with
\begin{equation}
D_{nl,nl'} = \left( \frac{Z_\mathrm{p}}{Z_\mathrm{t}} \right)^2 \frac{6n^2\max(l,\,l') 
\left[ n^2 - \max(l,\,l')^2 \right]}{2l+1}
\end{equation}
where $m$ and $Z_\mathrm{p}$ are the mass and charge of the projectile particle, 
$Z_\mathrm{t}$ is the charge of the target atom or ion, $\mu$ is the reduced mass 
of the colliding system, and $R_\mathrm{c}$ is the effective cut-off radius of 
the interaction at large impact parameters as given by \citet{1964PengellySeaton}.

In practice, $C_{nl,nl+1}$ is calculated using (\ref{eq:momcolrec2}) and 
$C_{nl,nl-1}$ is obtained using the principle of detailed balance
\begin{equation}
C_{nl,\,nl'}=\frac{2l'+1}{2l+1}C_{nl',\,nl}\,.
\label{eq:colmom detailed balance}
\end{equation}

%%%%%%%%%%%%%%%%%%%%%%%%%%%%%%%%%%%%%%%%%%%%%%%%%%

% Don't change these lines
\bsp	% typesetting comment
\label{lastpage}
\end{document}